\documentclass{ws-ijmpd}
\usepackage{epstopdf}
\usepackage{color}
\usepackage{rotating}

\begin{document}

\title{Cosmological models with squared trace in modified gravity}

\author{B. Mishra}
\address{Department of Mathematics, Birla Institute of Technology and Science-Pilani, Hyderabad Campus, Hyderabad 500078  India\\ bivudutta@yahoo.com}

\author{S. K. Tripathy}
\address{Department of Physics, Indira Gandhi Institute of Technology, Sarang, Dhenkanal, Odisha 759146, India\\ tripathy\_ sunil@rediffmail.com}

\author{Saibal Ray}
\address{Department of Physics, Government College of Engineering and Ceramic Technology, Kolkata 700010, West Bengal, India \& Department of Natural Sciences, Maulana Abul Kalam Azad University of Technology, Haringhata 741249, West Bengal, India\\
saibal@associates.iucaa.in}

\maketitle	
\begin{history}
\received{Day Month Year} \revised{Day Month Year}
\end{history}

\begin{abstract}
In this work we present a few simple cosmological models under the modified theory of gravity in the particular form of $f(R,\mathcal{T})=R+2f(\mathcal{T})$, where $R$ is the Ricci Scalar and $\mathcal{T}$ is the trace of the energy-momentum tensor. Two special cosmological models are studied with (i) hyperbolic scale factor and (ii) specific form of the Hubble parameter. The models are observed to predict relevant cosmological parameters closer to the observational values. Both the models reduce to overlap with the $\Lambda$CDM model at late times. We have discussed some interesting results related to wormhole solutions as evolved from our model. The possible occurrence of Big Trip in wormholes for the models are also discussed.
\end{abstract}

\keywords{modified gravity; cosmological model; wormhole solution}

\section{Introduction} 
The need of modified/alternative/extended theories of gravity is to answer certain shortcomings in the standard cosmology under the framework of General Relativity (GR). Basically, various notable cosmological observations, viz. Cosmic Microwave Background Radiation (CMBR)~\cite{Spergel2003,Spergel2007}, high redshift supernovae~\cite{Riess1998}, baryon acoustic oscillations~\cite{Percival2010}, Planck data~\cite{Ade2014} and supernovae of type Ia~\cite{Perlmutter1999,Bennett2003}, lead to the obvious conclusion that our universe along with its expanding phase also accompanied by the late-time cosmic accelerated expansion. It is therefore argued in Ref.~\cite{Deb2019a} that though till now GR seems to be the sublime tool to study the large-scale structure of the universe successfully, but in connection to the late time acceleration of the universe has been facing a major set back. However, the Einstein field equations with the cosmological constant~\cite{Sahni2000,Peebles2003} along with further mysterious assumption of {\it Dark Matter} (DM)~\cite{Overduin2004,Baer2015} and {\it Dark Energy} (DE) could explain the accelerated expansion of the universe.\\

In this critical situation several scientists became motivated to develop alternative gravity theories (without considering hypothetical DM and DE components) through extensions of the Einstein-Hilbert action. Among these proposed theories, which provide a deeper and wider understanding of the quantum mechanical as well as gravitational field theory at the high energy densities, some notable are as follows: the effective first-order approximation to quantum gravity~\cite{Buchbinder1992,Parker2009}, modifying the Lagrangian density via a simple function $f\left(R\right)$ in the Einstein-Hilbert action, known as $f(R)$ gravity~\cite{Capozziello2002,Nojiri2003,Carroll2004,Bertolami2007}, $f\left(G\right)$ gravity~\cite{Bamba2010,Rodrigues2014}, $f\left(R,G\right)$ gravity~\cite{Nojiri2005a}, $f\left(\mathbb{T}\right)$ gravity~\cite{Bengochea2009,Linder2010,Bohmer2011}, Brans-Dicke (BD) gravity~\cite{Avilez2014,Bhattacharya2015}. Here the symbols $R$, $G$ and $\mathbb{T}$ are respectively the Ricci scalar, Gauss-Bonnet scalar and torsion scalar.\\

However, among the above all, the recent promising alternative theory is the $f(R,\mathcal{T})$ theory~\cite{Harko2011} which is the generalization of $f(R)$ theory~\cite{Nojiri2007,Nojiri2009}, $\mathcal{T}$ being the trace of the energy-momentum tensor. In this modified theory, several cosmological models have been studied by scientists through the last decade. Under this $f(R,\mathcal{T})$ gravity, Harko et al.~\cite{Harko2014} provided an open irreversible thermodynamic interpretation of the cosmological model. Shabani and Farhoudi~\cite{Shabani2014} studied several cosmological parameters to obtain the most acceptable cosmological results whereas Shamir~\cite{Shamir2015} has explored the exact solution in Bianchi I space-time. Under the formalism with $f(R,\mathcal{T})$ gravity, Moraes and Correa~\cite{Moraes2016a} obtained  the cosmological parameters and interestingly their results are in nice agreement with the recent observational constraints. Mishra et al.~\cite{Mishra2016} have investigated the dynamics of an anisotropic universe by using a rescaled functional $f(R,\mathcal{T})$ and generated the idea of a time variable cosmological constant. In this modified gravity, Mishra and Vadrevu~\cite{Mishra2017} have obtained and analyzed the solution with the quadratic form of Ricci Scalar in Einstein-Rosen spacetime. 

A mathematical formalism have been developed in this theory by Mishra et al.~\cite{Mishra2018a,Mishra2018b} and observed that increase in cosmic anisotropy affects the energy conditions. They~\cite{Mishra2018c} have also noticed that the models remain in the quintessence phase from an anisotropic spacetime when the matter acts as viscous fluid. A new scale factor have been introduced by Mishra et al.~\cite{Mishra2018d}, which consist of two factors, each of them dominating at an early and late cosmic epochs, respectively. Tarai and Mishra~\cite{Tarai2018} investigated the cosmological model in an anistropic spacetime with electromagnetic field whereas Esmaeili and Mishra~\cite{Esmaeili2018} have introduced the hyperbolic scale factor to investigate the behaviour of the model. Several works are also done on the bouncing and other cosmology in $f(R,\mathcal{T})$ gravity~\cite{Shabani2018,Singh2018,Tripathy2019,Tripathy2020a,Tripathy2020b,Mishra2020d}. It is to note that, plenty of works under $f(R,\mathcal{T})$ gravity in the field of astrophysics are available in literature~\cite{Azizi2013,Sharif2014,Alhamzawi2015,Moraes2016,Das2016,Das2017,Moraes2017,Deb2018a,Deb2018b,Deb2019a,Deb2019b,Biswas2019,Biswas2020,Ray2020}

Motivated by the above background, we present a few simple cosmological models under the modified theory of gravity  along with some interesting results related to wormhole solutions. Following is the scheme of our work: the mathematical formalism for the cosmological system in Sec.~2. In Sec.~3 we have provided physical parameters of the cosmological models along with two specific models are dealt with under two specific cases, viz. (i) model with Hyperbolic Scale Factor and (ii) model with specific form of Hubble Parameter. The wormhole solutions and Big Trip case study have been done in Sec.~4. The last Sec.~5 is kept for some concluding remarks.

\section{Basic mathematical formalism}
Considering the matter-geometry coupling, the action for a geometrically modified theory is given by 
\begin{equation} \label{eq:1}
S=\frac{1}{16\pi}\int d^4x\sqrt{-g} f(R,\mathcal{T})+ \int d^4x\sqrt{-g}\mathcal{L}_m,
\end{equation}
where the function defined in the action $S$ is an arbitrary function of the Ricci scalar $R$ and trace of the energy-momentum tensor $\mathcal{T}$ along with $\mathcal{L}_m$ as the matter Lagrangian and $g$ as the metric tensor. 

To fulfill the requirement of any modified theory of gravity one needs to modify GR either via the geometry part or the matter side. The matter field with a positive energy density and negative pressure generally used as extra terms to conceive the cosmic acceleration~\cite{Mishra2020d}. According to the proposal of Harko et al.~\cite{Harko2011}, the functional form which governs the matter-curvature coupling is arbitrary, so that one can have different choices of the functional $f(R,\mathcal{T})$ to generate different kind of cosmological models. They have suggested three functional ways of coupling scheme, viz., (i) $f(R,\mathcal{T})=R+2f(\mathcal{T})$, (ii) $f(R,\mathcal{T})=f_1 (R)+f_2 (\mathcal{T})$ and (iii) $f(R,\mathcal{T})=f_1 (R)+f_2 (R)f_3 (\mathcal{T})$, where $f_1 (R)$, $f_2 (R)$, $f_2 (\mathcal{T})$, $f_3 (\mathcal{T})$ are arbitrary functions of their respective arguments. However, out of these three types, the scheme $f(R,\mathcal{T})=R+2f(\mathcal{T})$ has been used by several investigators~\cite{Reddy2013,Moraes2014,Kumar2015,Shamir2015,Moraes2016b,Tarai2018,Aygun2019,Mishra2020d} and hence we have adopted the assumption $f(R,\mathcal{T})=f_1 (R)+f_2 (\mathcal{T})$ in our present investigation.

Now, under the above splitting scheme, the action for a minimal matter-geometry coupling becomes
\begin{equation} \label{eq:2}
S=\frac{1}{16\pi}\int d^4x\sqrt{-g} f_1(R)+\frac{1}{16\pi}\int d^4x\sqrt{-g}f_2(\mathcal{T})+ \int d^4x\sqrt{-g}\mathcal{L}_m.
\end{equation}

It can be seen that (i) when the middle term of Eq.~(\ref{eq:2}) vanishes, and (ii) in the first term $f_1(R)=R$, then the action \ref{eq:2} reduces to be the action of General Relativity (GR). We have used the geometrized unit as $G=1=c$. 

Now, varying the action with respect to the metric $g_{\mu\nu}$, the modified field equation can be obtained as
\begin{eqnarray} \label{eq:3}
&\qquad\hspace{-20.5cm} R_{\mu\nu}-\frac{1}{2}f^{-1}_{1,R} (R)f_1(R)g_{\mu\nu}=\nonumber \\ f^{-1}_{1,R}(R)\left[\left(\nabla_{\mu} \nabla_{\nu}-g_{\mu\nu}\Box\right)f_{1,R}(R)+\left[8\pi +f_{2,T}(\mathcal{T})\right]T_{\mu\nu} +\left[f_{2,\mathcal{T}}(\mathcal{T})p+\frac{1}{2}f_2(\mathcal{T})\right]g_{\mu\nu}\right],
\end{eqnarray}
where $\mathcal{L}_m=-p$, is the pressure of the cosmic fluid and the other mathematical notations are as follows: $f_{1,R} (R)\equiv \frac{\partial f_1(R)}{\partial R},~f_{2,\mathcal{T}} (\mathcal{T})\equiv \frac{\partial f_2(\mathcal{T})}{\partial \mathcal{T}},~f^{-1}_{1,R} (R) \equiv \frac{1}{f_{1,R} (R)}$ whereas 
$T_{\mu\nu}=-\frac{2}{\sqrt{-g}}\frac{\delta\left(\sqrt{-g}\mathcal{L}_m\right)}{\delta g^{\mu\nu}}$ serves as the energy-momentum tensor related to the matter Lagrangian.

Let us now construct, from the above field equation (\ref{eq:3}), a GR-like modified gravity theory, which yields 
\begin{equation}\label{eq:7}
G_{\mu\nu}= \left[8\pi +f_{2,\mathcal{T}}(T)\right]T_{\mu\nu}+\left[f_{2,\mathcal{T}}(\mathcal{T})p+\frac{1}{2}f_2(\mathcal{T})\right]g_{\mu\nu}
= \kappa_{\mathcal{T}}\left[T_{\mu\nu}+ T^{int}_{\mu\nu}\right],
\end{equation}
where $\kappa_{T}= 8\pi +f_{2,\mathcal{T}}(\mathcal{T})$ is the redefined Einstein constant. It is to note that (i) $f_{2,\mathcal{T}}(\mathcal{T})$ and consequently $\kappa_{\mathcal{T}}$  become constants for a linear  functional $f_2(\mathcal{T})$ and (ii) $\kappa_{\mathcal{T}}$ evolves with time and dynamically mediates the coupling between the geometry and matter for any non-linear choices of the functional $f_{2}(\mathcal{T})$. 

Hence from Eq. (\ref{eq:7}), the effective energy-momentum tensor is given by
\begin{equation}\label{eq:8}
T^{int}_{\mu\nu}=\left[ \frac{f_{2,\mathcal{T}}(\mathcal{T})p+\frac{1}{2}f_2(\mathcal{T})}{8\pi +f_{2,\mathcal{T}}(\mathcal{T})}\right]g_{\mu\nu}.
\end{equation}

It is to note that if we drop the $\mathcal{T}$ dependent part of the functional $f(R,\mathcal{T})=R+f_2(\mathcal{T})$, then the interactive contribution to the energy-momentum tensor does vanish. It is argued~\cite{Tripathy2020a} that a minimal matter-geometry coupling in the action behaves like an extra matter field that is responsible to provide an acceleration as derived from quantum effects due to energy fluctuation and this might have a leading role for non-vanishing divergence of the energy-momentum tensor $T_{\mu\nu}$. However, different functional forms of $f(R,\mathcal{T})$ containing non-linear terms in $R$, it is possible to have a fluid acceleration even when the coupling constant $\lambda$ vanishes. In that case, the model reduces to the usual $f(R)$ class of gravity model. Our interest in the present work is to obtain extended gravity models with a minimum number of adjustable parameters which under suitable conditions should reduce to GR and therefore, we consider $f(R)=R$.

Therefore, we are keeping this in mind that a suitable choice of the functional $f_2(\mathcal{T})$ may provide a plausible cosmological model which will be in conformation with the accelerating phase of the present universe. This is the key point for our motivation to investigate some hyperbolic scale factor based cosmological models in the theory of modified gravity. To proceed on we consider the form of $f_2(\mathcal{T})$ as
\begin{equation}\label{eq:9}
\frac{1}{2}f_2(\mathcal{T})=\lambda \mathcal{T}^2 +\Lambda_0.
\end{equation}

This immediately provides
\begin{eqnarray}
%\kappa_T &=& 8\pi+4\lambda \mathcal{T} ,\label{eq:10}\\
T^{int}_{\mu\nu} &=& \frac{g_{\mu\nu}}{\kappa_\mathcal{T}}\left[ \mathcal{T}\left(4 p+\mathcal{T}\right)\lambda+\Lambda_0\right].\label{eq:11}
\end{eqnarray}

It is interesting to observe that for $\lambda=0$, one can get $T^{int}_{\mu\nu} = g_{\mu\nu}\frac{\Lambda_0}{8\pi}$ which implies that the late time acceleration is governed by the constant $\Lambda_0$ and therefore shows behavior of the erstwhile cosmological constant as envisioned by Einstein in GR.

\section{Cosmological models under the $f(R,\mathcal{T})$ gravity}
Let us now consider the flat FRW spacetime
\begin{equation}\label{eq:12}
ds^2 = dt^2 - a^2(t)(dx^2+ dy^2+dz^2),
\end{equation}
with $a(t)$ as its scale factor. 

As a specific assumption we also consider the universe to be filled with perfect fluid so that the energy-momentum tensor can be provided in the form
\begin{equation}\label{eq:13}
T_{\mu\nu}=(p+\rho)u_{\mu}u_{\nu} - pg_{\mu\nu},
\end{equation}
where $p$ and $\rho$ represent respectively the pressure and energy density of the cosmic fluid whereas $u^{\mu}u_{\mu}=1$ with $u^{\mu}$ as the velocity of the comoving coordinate system.

Therefore, the Einstein field equations in the modified gravity theory for a flat FRW spacetime can be obtained as
\begin{eqnarray}
2\dot{H}+3H^2 &=& -(8\pi+3\lambda) p + \lambda \rho+\Lambda_0, \label{eq:15}\\
3 H^2 &=& (8\pi+3\lambda) \rho - \lambda p +\Lambda_0, \label{eq:16}
\end{eqnarray}
where the ordinary time derivatives are presented as overhead dots. 

Hence, we shall derive the dynamical parameters for the cosmological models where the pressure and energy density can be obtained as
\begin{eqnarray}
p&=&-\frac{1}{\kappa_\mathcal{T}(\kappa_\mathcal{T}+2\lambda)}\left[ 2(8\pi+3\lambda)\dot{H}+3(8\pi+3\lambda)H^2\right]+\frac{\Lambda_0}{\kappa_\mathcal{T} +2\lambda}, \label{eq:17}\\
\rho &=& \frac{1}{\kappa_\mathcal{T}(\kappa_\mathcal{T}+2\lambda)}\left[ -2\lambda\dot{H}+3(8\pi+3\lambda)H^2\right]-\frac{\Lambda_0}{\kappa_\mathcal{T} +2\lambda}. \label{eq:18}
\end{eqnarray}

Now, a common concept of the perfect fluid Equation of State (EOS) frequently used in cosmology as well as astrophysics is characterized by a dimensionless number in the form $\omega=\frac{p}{\rho}$ to understand the physical feature of the system under consideration. Therefore, in the proposed model we get the EOS parameter $\omega$ in the following form
\begin{equation}\label{eq:19}
\omega=-1+4(2\pi+\lambda)\left[\frac{2\dot{H}}{2\lambda\dot{H}-3(8\pi+3\lambda)H^2+ \kappa_{T} \Lambda_0}\right].
\end{equation}

It has been argued by Mishra et al.~\cite{Mishra2020d} that the energy conditions put some additional constraints on any model as far as viable physical feature is concerned, specifically the acceleration or deceleration of cosmic fluid and hence emergence of singularity, known as Big Rip.  In different contexts these type of elegant conditions are employed to derive general results which have been successful in varieties of situation~\cite{Alam2004,Perez2006,Szvyd2006,Sharif13,Capozziello18}. Hence, we are interested to find out the energy conditions, specifically the null energy condition (NEC), $\rho+p \geq 0$, which can be provided as 
\begin{equation}\label{eq:20}
p+\rho=-4(2\pi+\lambda)\left[\frac{2\dot{H}}{\kappa_\mathcal{T}(\kappa_\mathcal{T}+2\lambda)}\right].
\end{equation}

Let us consider the following two specific cosmological models to study in the framework of the modified $f(R,\mathcal{T})$ theory of gravity.

\subsection{Model with Hyperbolic Scale Factor}
In this subcase, we are interested to study the model of the universe with hyperbolic scale factor $a(t)=sinh^{\frac{1}{m}}(\beta t)$ in a modified gravity~\cite{Amirhashchi11,Chawla13}. The Hubble parameter for such a hyperbolic scale factor  can be expressed as 
\begin{equation}\label{eq:21}
H=\frac{\beta}{m}coth(\beta t),
\end{equation}
where $\beta$ and $m$ are positive constant parameters. It is worth to mention here that, the $\Lambda$CDM model suffers from the well known $H_0$ tension. The $H_0$ tension arises due to the discrepancies in the values of the Hubble parameter obtained in the local distance ladder measurement~\cite{Reiss2016,tension} and  an indirect measurement from the CMB temperature from Planck collaboration~\cite{Aghanim2018}. While the local distance ladder measurement of Reiss~\cite{Reiss2016,tension} yields $H_0=74.03\pm 1.42~km~s^{-1}Mpc^{-1}$, the CMB temperature measurement~\cite{Aghanim2018} provides $H_0=67.36\pm 0.54~km~s^{-1}Mpc^{-1}$. Other recent measurements include that of the SH$_o$ES~\cite{Reid2019}, $H_0=73.5\pm 1.4~km~s^{-1}Mpc^{-1}$ and that of the H0LiCOW collaboration~\cite{Wong2019}, $H_0=73.3^{+1.7}_{-1.8}~km~s^{-1}Mpc^{-1}$. The $H_0$ tension may hint for a new Physics involving the dynamics of dark matter/dark energy. Also, it may hint for a possible departure from  $\Lambda$CDM model. In the present work, we wish to adjust the parameters of the hyperbolic scale factor in such a manner to obtain the Hubble parameter at the present epoch ($t\approx 13.8$ Gyr) close to the observational values as mentioned above. In view of this, we consider, $m=0.057$ and three representative values of $\beta$ namely $\beta=4.25,3.95$ and $3.65$. In Fig.1, we show the Hubble parameter for the hyperbolic scale factor for the representative values of $\beta$. These values of the parameter $\beta$ predicts the Hubble parameter at the present epoch respectively as $H_0=74.73, 69.29$ and $64.64$~$km~s^{-1}Mpc^{-1}$.
\begin{figure}[h!]
\centering
\includegraphics[width=8cm]{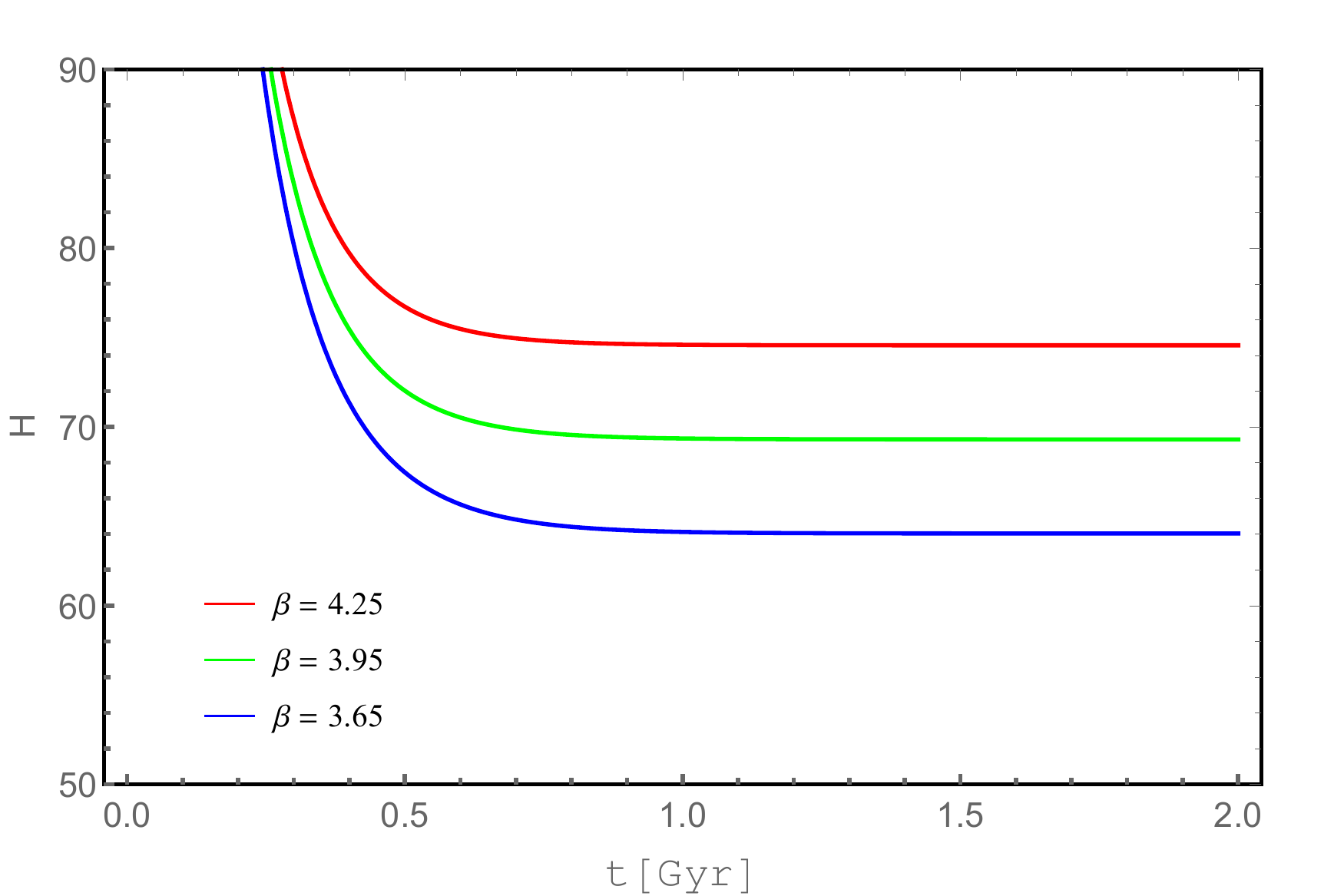}
\caption{Hubble parameter corresponding to the hyperbolic scale factor (for Eq. (\ref{eq:21}))}
\end{figure}
%%%%%%%%%%%%%%%%%%%%%%%%%%%%%%%%%%%%%%%%%%%

The deceleration parameter and the jerk parameter for this scale factor can be respectively found out as 
\begin{equation}\label{eq:22a}
q=-1 +\frac{d}{dt}(\frac{1}{H}) =-1+m sech^2(\beta t), 
\end{equation}

\begin{equation}\label{eq:22b}
j= {\frac{\ddot H}{H^3}} - (2 + 3q)= 1+m(2m-3)sech^2\beta t.
\end{equation}

It is observed from the above equations that both the deceleration parameter as well as the jerk parameter are dynamical quantities and  asymptotically approach to $-1$ and $1$ respectively. These two geometrical parameters are shown as function of time respectively in Figs. 2 and 3. During the evolution, while $q$ remains in the negative domain, the jerk parameter remains in the positive domain.  For a given value of $\beta$, $q$ decreases initially and reaches to its asymptotic value at late times. The jerk parameter on the other hand, increases initially and reaches its asymptotic value at late times. The choice of the parameter $\beta$ affects these quantities during an initial epoch. At an initial epoch, higher is the value of $\beta$, lower is the value of $q$ and higher is the value of $j$. However at a late epoch, the behaviour of the deceleration parameter and the jerk parameter remain unaffected by the choice of $\beta$. At the present epoch ($t\approx 13.8$ Gyr), the deceleration parameter becomes $-1$ and the jerk parameter becomes $1$.
%%%%%%%%%%%%%%%%%%%%%%%%%%%%%%%%%%%%%%%%%%%
\begin{figure}[h!]
\centering
\includegraphics[width=8cm]{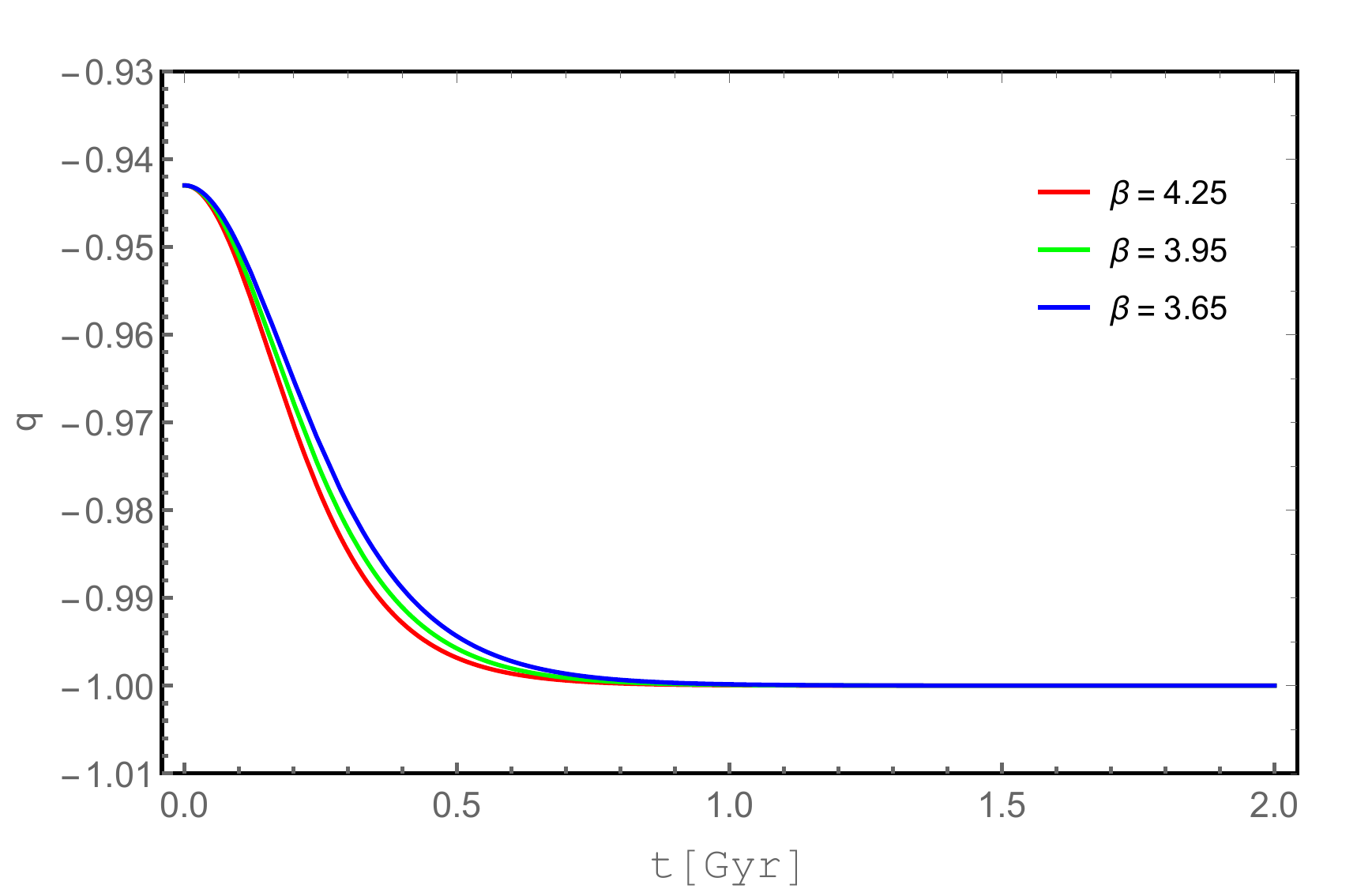}
\caption{Plot for variation of deceleration parameter $q$ w.r.t. time $t$ (for Eq. (\ref{eq:22a}))}
\end{figure}
%%%%%%%%%%%%%%%%%%%%%%%%%%%%%%%%%%%%%%%%%%%

%%%%%%%%%%%%%%%%%%%%%%%%%%%%%%%%%%%%%%%%%%%
\begin{figure}[h!]
\centering
\includegraphics[width=8cm]{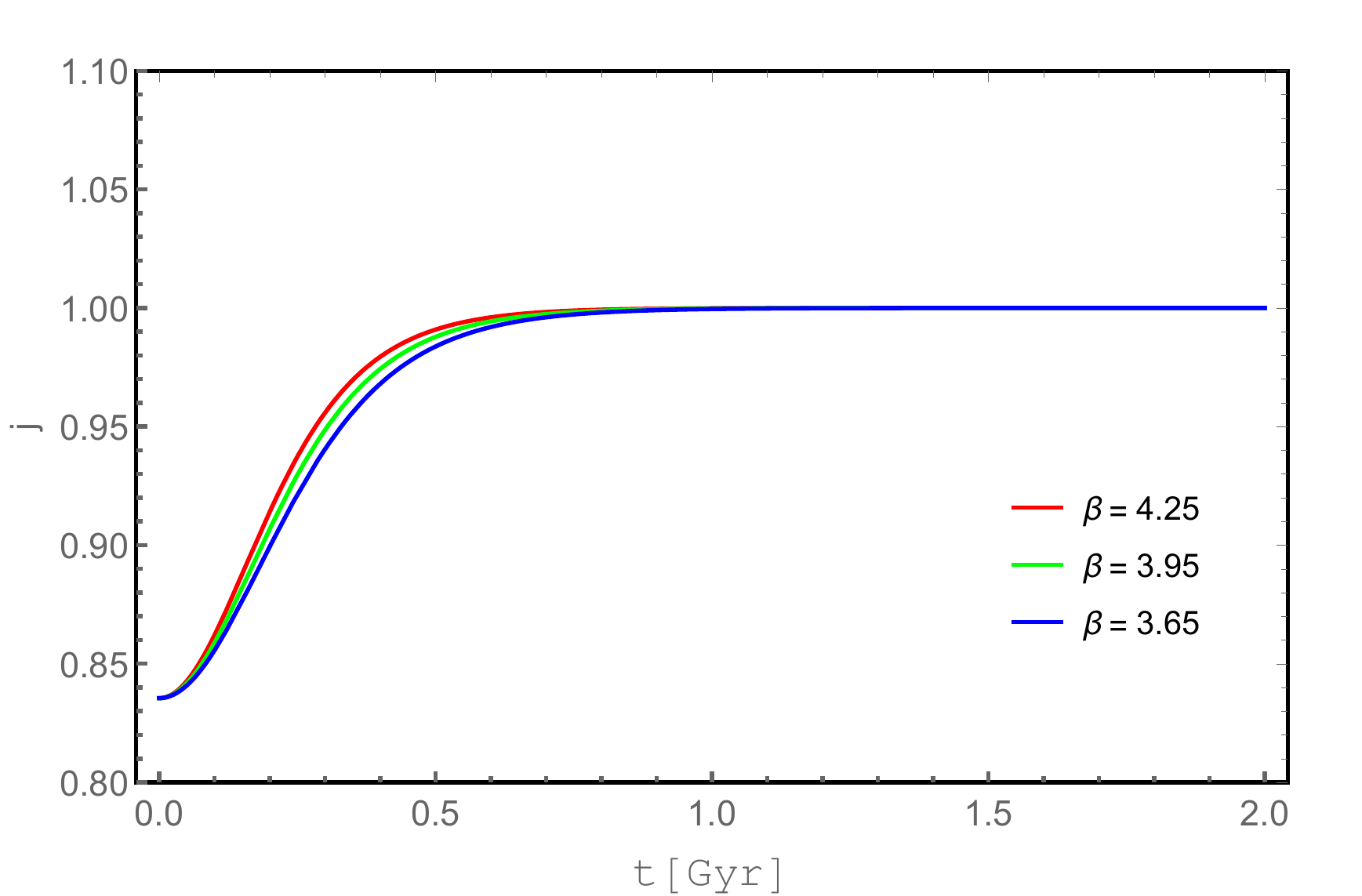}
\caption{Plot for variation of jerk parameter $j$ w.r.t. time $t$ (for Eq. (\ref{eq:22b}))}
\end{figure}
%%%%%%%%%%%%%%%%%%%%%%%%%%%%%%%%%%%%%%%%%%%

With the hyperbolic scale factor, we get the pressure, energy density and EOS parameter as follows:
\begin{eqnarray}
p&=&-\frac{\beta^2}{m^2\kappa_\mathcal{T}(\kappa_\mathcal{T}+2\lambda)}\left[- 2m(8\pi+3\lambda) cosech^2(\beta t)+3(8\pi+3\lambda)coth^2(\beta t)\right] \nonumber \\ 
  &+& \frac{\Lambda_0}{\kappa_\mathcal{T} +2\lambda}, \label{eq:22}\\
\rho &=& \frac{\beta^2}{m^2\kappa_\mathcal{T}(\kappa_\mathcal{T}+2\lambda)}\left[ 2m\lambda  coseh^2(\beta t)+3(8\pi+3\lambda)coth^2(\beta t)\right] \nonumber \\ 
     &-& \frac{\Lambda_0}{\kappa_\mathcal{T} +2\lambda}, \label{eq:23}\\
\omega&=&-1+4(2\pi+\lambda)\frac{2m cosech^2(\beta t)}{2m\lambda cosech^2{\beta t}+3(8\pi+3\lambda)coth^2 (\beta t)+m\kappa_\mathcal{T}\Lambda_0}.
\end{eqnarray}

In Fig. 4, the dynamical behaviour of the equation of state parameter for the hyperbolic scale factor model is shown. We use the three representative values of $\beta$ to plot the figures. Also, we consider the coupling constant $\lambda=-0.5$ to ensure that, the energy density for the present model remains positive throughout the cosmic evolution. For plotting the figures we choose the parameter $\Lambda_0$ to be $0.1$. The equation of state parameter dynamically evolves from a low negative value at an early epoch to asymptotically become $-1$ at late times. The choice of the parameter $\beta$ affects $\omega$ in an initial epoch in the sense that, higher the value of $\beta$, higher is the value of $\omega$. At the present epoch, the hyperbolic scale factor model predicts $\omega=-1$.
%%%%%%%%%%%%%%%%%%%%%%%%%%%%%%%%%%%%%%
\begin{figure}
\centering
\includegraphics[width=8cm]{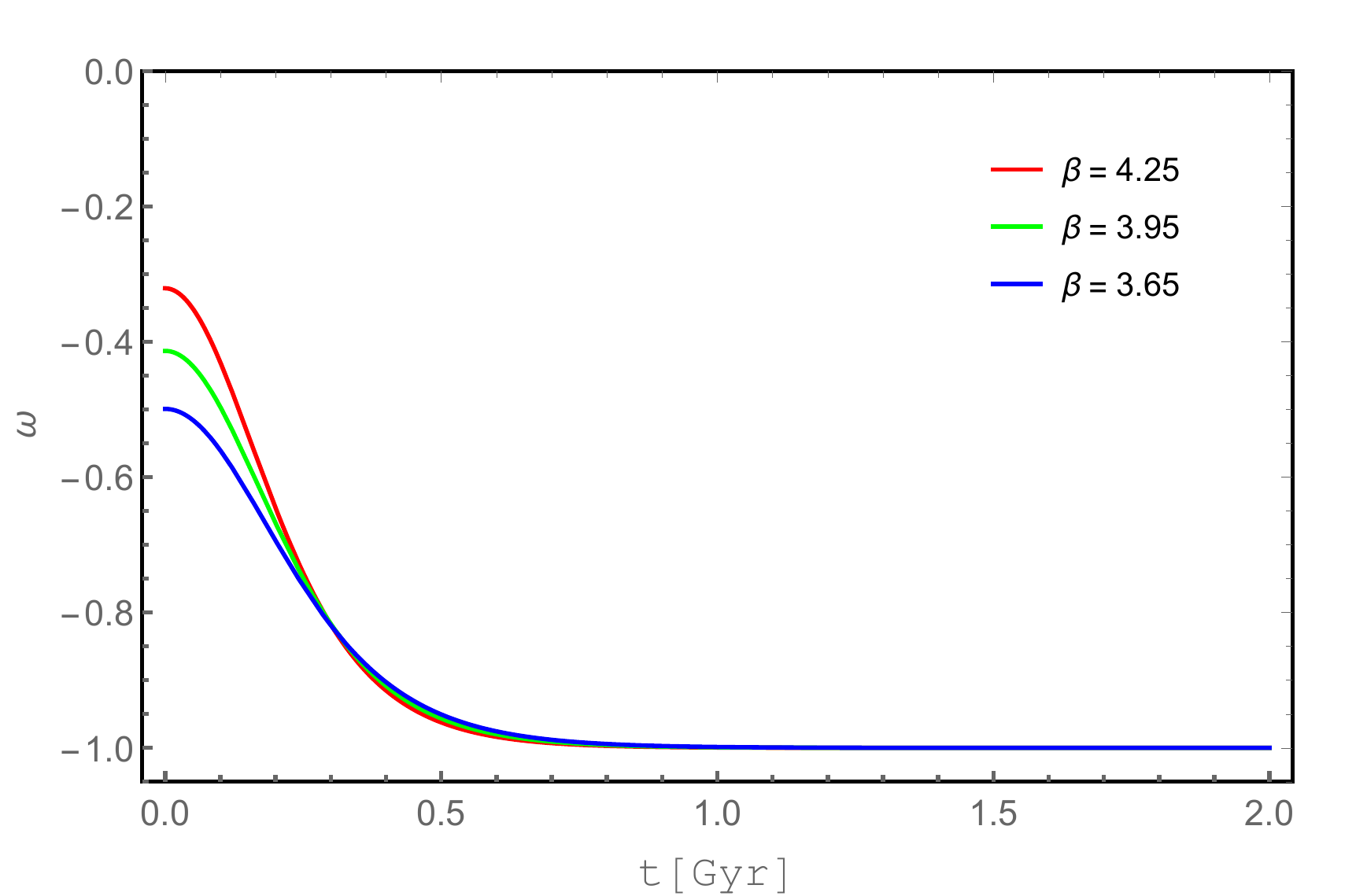}
\caption{Plot for variation of EOS parameter $\omega$ w.r.t. time $t$ (for the hyperbolic model)}
\end{figure}
%%%%%%%%%%%%%%%%%%%%%%%%%%%%%%%%%%%%%%%

\begin{figure}[h!]
\centering
\includegraphics[width=8cm]{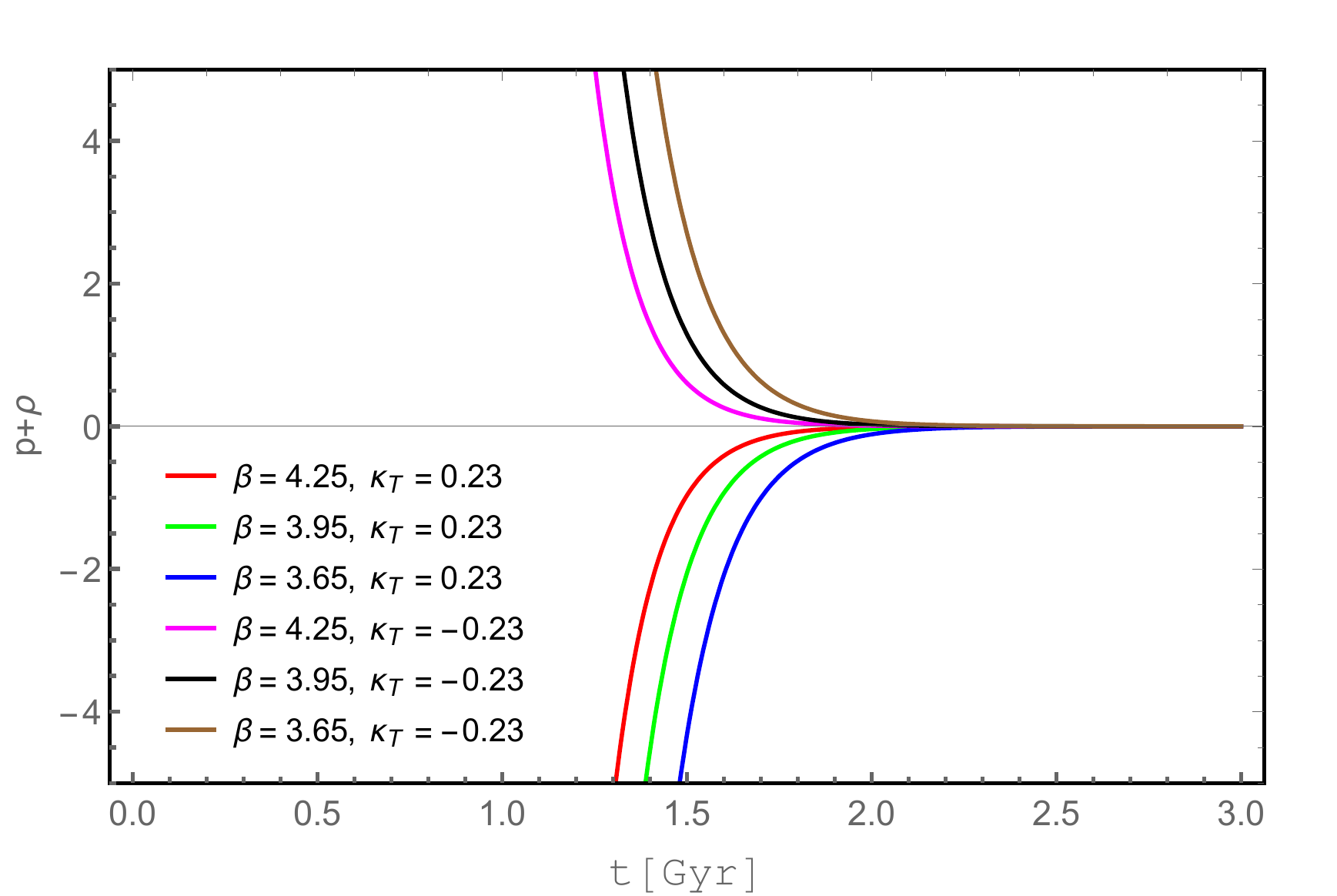}
\caption{p+$\rho$ vs time for the hyperbolic scale factor model. The upper panel is for negative value of $\kappa_T$ and the lower panel is for positive $\kappa_T$.}
\end{figure}
%%%%%%%%%%%%%%%%%%%%%%%%%%%%%%%%%%%%%%%%%%%

We obtain the null energy condition (NEC) for the hyperbolic scale factor as
\begin{equation} \label{eq:24}
p+\rho=\frac{8(2\pi+\lambda)\beta^2}{m\kappa_\mathcal{T}(\kappa_\mathcal{T}+2\lambda)}cosech^2\beta t.
\end{equation}

In Fig. 5, the null energy condition is shown for the three representative values of $\beta$ and $\lambda=-0.5$. Here we have used two different values of $\kappa_T$, i.e. $\kappa_\mathcal{T}=0.23$ and $\kappa_\mathcal{T}=-0.23$. While for positive value of $\kappa_\mathcal{T}$, all the models with different $\beta$, violate the NEC, however it is satisfied for negative value of $\kappa_\mathcal{T}$.

\subsection{Model with specific form of Hubble parameter}
In this subcase, we consider a specific form of the Hubble parameter~\cite{Banerjee05}, which can be obtained as 
\begin{equation} \label{eq:25}
H=\frac{\alpha e^{\alpha nt}}{e^{\alpha nt}-1},
\end{equation}
where $\alpha>0$, $n>0$ are positive constants and the corresponding scale factor can be calculated as $a(t)=(e^{\alpha nt}-1)^{\frac{1}{n}}$.  As in the previous model, in this model also, we have adjusted the parameters of the scale factor so that, we obtain a reasonable value of the Hubble parameter at the present epoch. We consider $\alpha=74.25$ and three different values of $n$ namely $n=0.15, 0.20$ and $0.25$. In Fig.2, the Hubble parameter is shown for the representative values of $n$. It can be observed from the figure that, all the models coincide to the same curve at a late epoch. This is because of the presence of the exponential factor in the scale factor. All the models predict the same value of the Hubble parameter at the present epoch as $H_0=74.25 ~km~s^{-1}Mpc^{-1}$. 

\begin{figure}[h!]
\centering
\includegraphics[width=8cm]{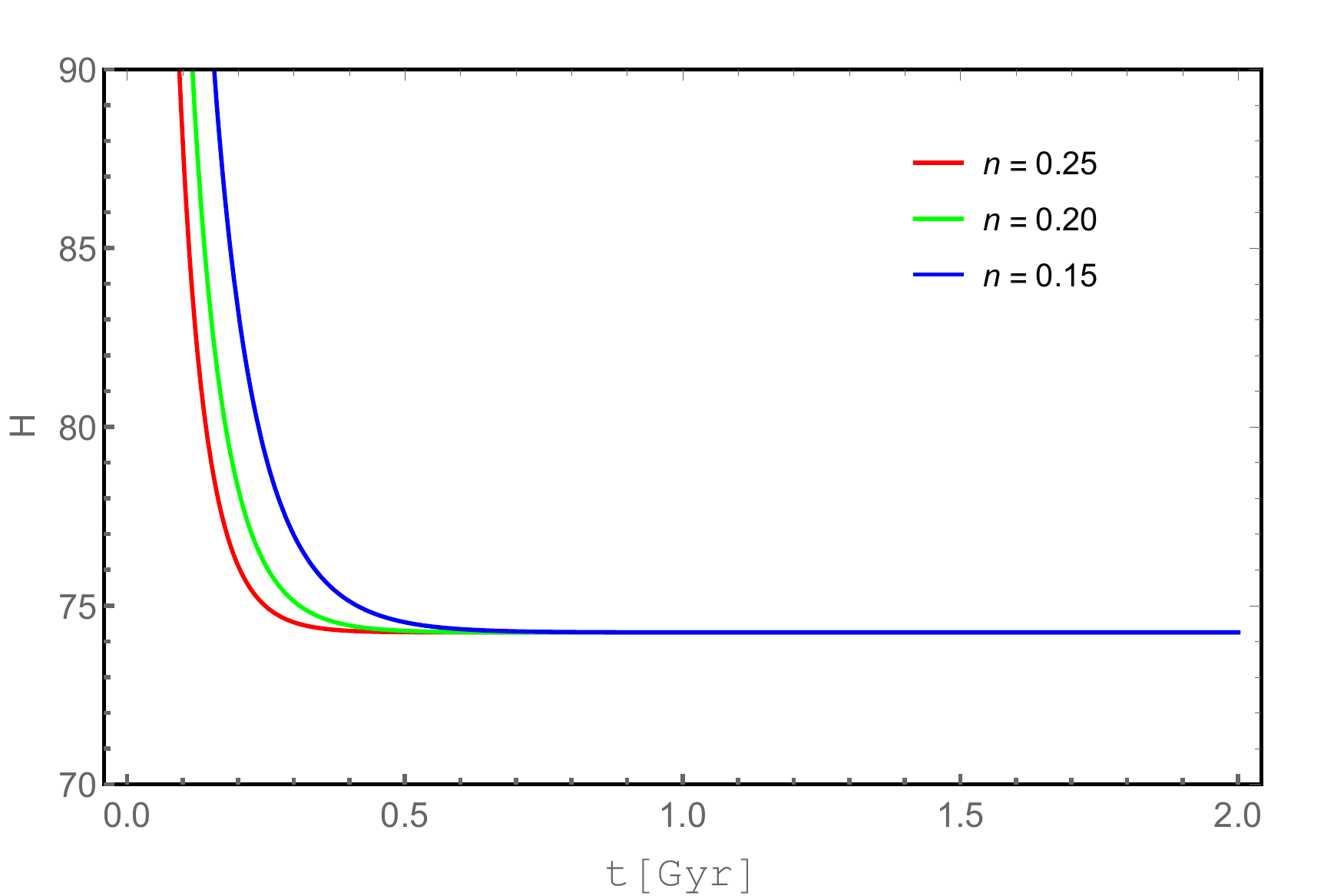}
\caption{Hubble parameter corresponding to the specific form of the Hubble parameter (for Eq. (\ref{eq:25}))}
\end{figure}
%%%%%%%%%%%%%%%%%%%%%%%%%%%%%%%%%%%%%%%%%%%

The deceleration parameter and jerk parameter can be expressed as
\begin{equation} \label{eq:26a}
q= -1+\frac{n}{e^{\alpha n t}} 
\end{equation}

\begin{equation} \label{eq:26b}
j= 1-(n+1)\left(\frac{n}{e^{\alpha nt}}\right)+\left(\frac{n}{e^{\alpha nt}}\right)^2.
\end{equation}

%%%%%%%%%%%%%%%%%%%%%%%%%%%%%%%%%%%%%%%%%%%
\begin{figure}[h!]
\centering
\includegraphics[width=8cm]{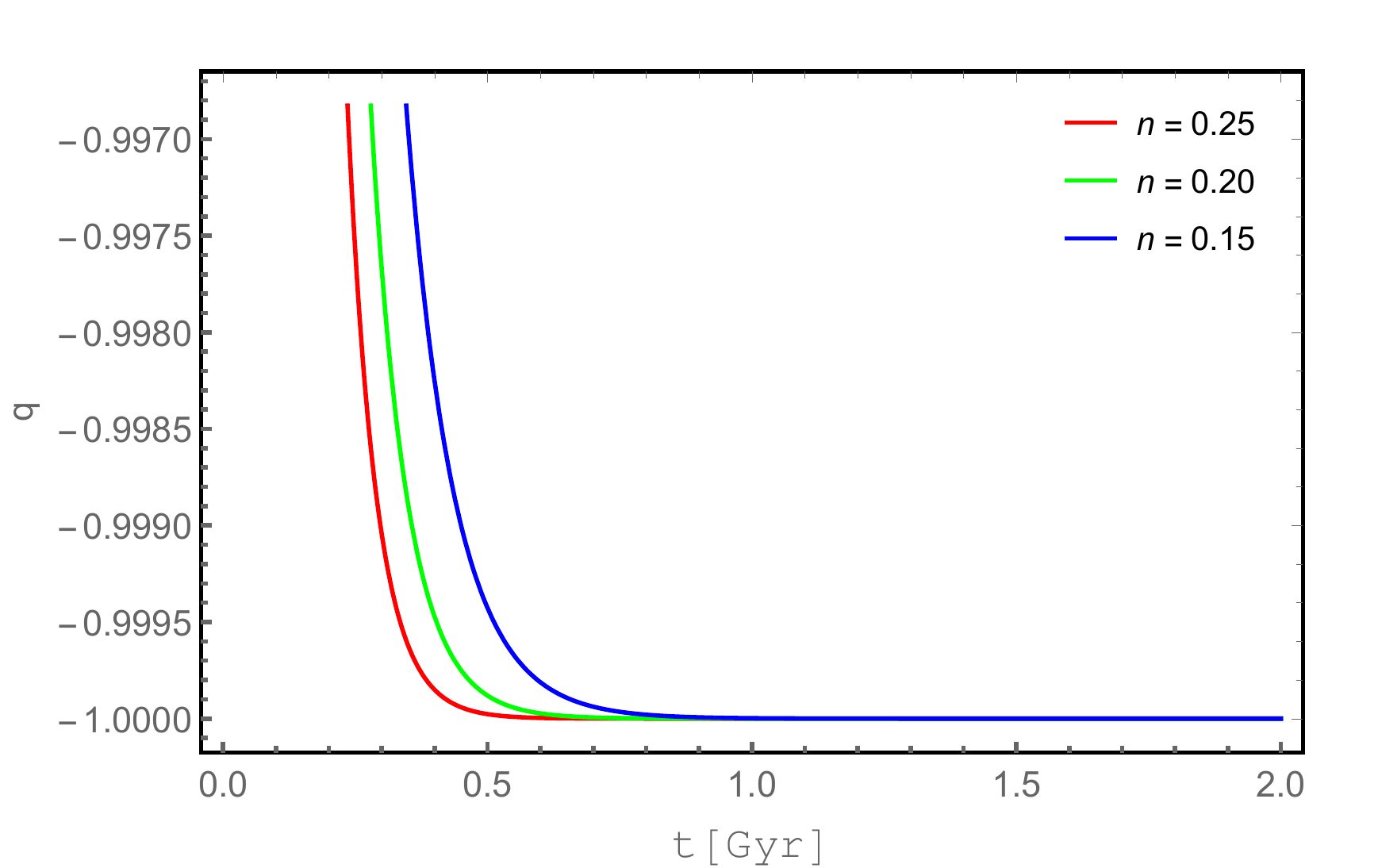}
\caption{Plot for variation of deceleration parameter $q$ w.r.t. time $t$ (for Eq. (\ref{eq:26a}))}
\end{figure}
%%%%%%%%%%%%%%%%%%%%%%%%%%%%%%%%%%%%%%%%%%%

%%%%%%%%%%%%%%%%%%%%%%%%%%%%%%%%%%%%%%%%%%%
\begin{figure}[h!]
\centering
\includegraphics[width=8cm]{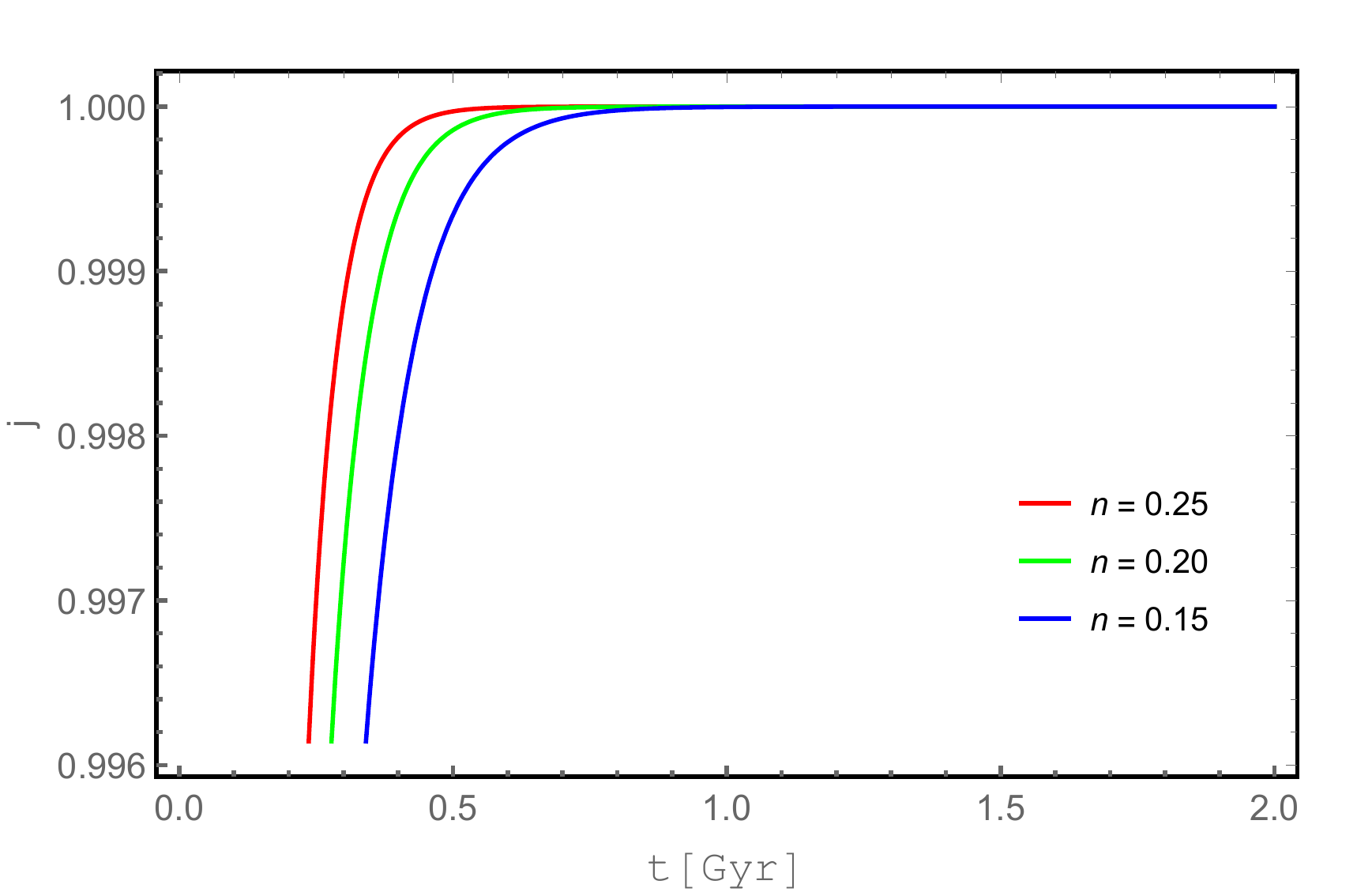}
\caption{Plot for variation of jerk parameter $j$ w.r.t. time $t$ (for Eq. (\ref{eq:26b}))}
\end{figure}
%%%%%%%%%%%%%%%%%%%%%%%%%%%%%%%%%%%%%%%%%%%

In Fig. 7 we show the variation of the deceleration parameter and in Fig. 8, the evolution of the jerk parameter is shown. We consider three different values of the parameter $n$. Since the Hubble parameter contains an exponential factor, the deceleration parameter quickly decreases from a low negative value to become $-1$. The parameter $n$ affects the rate of decrement of $q$. Lower the value of $n$, higher the rate of decrement in $q$. The model predicts  $q=-1$ at the present epoch. The jerk parameter is a positive quantity and increases to become $1$. The rate of increment in $j$ is decided by the parameter $n$ in the sense that, higher the $n$, higher is the increment rate.

With this scale factor, we can obtain the pressure, energy density, EOS parameter as
\begin{eqnarray}
p&=&-\frac{\alpha^2 e^{\alpha nt}}{\kappa_\mathcal{T}(\kappa_\mathcal{T}+2\lambda)(e^{\alpha nt}-1)^2}\left[-2n(8\pi+3\lambda)+3(8\pi+3\lambda) e^{\alpha nt}\right]+\frac{\Lambda_0}{\kappa_\mathcal{T} +2\lambda}\label{eq:26}\\
\rho &=& \frac{\alpha^2 e^{\alpha nt}}{\kappa_\mathcal{T}(\kappa_\mathcal{T}+2\lambda)(e^{\alpha nt}-1)^2}\left[ 2\lambda n+3(8\pi+3\lambda)e^{\alpha nt}\right]-\frac{\Lambda_0}{\kappa_\mathcal{T} +2\lambda}\label{eq:27}\\
\omega&=&-1+8(2\pi+\lambda)\frac{n}{2\lambda n+3(8\pi+3\lambda)e^{\alpha n t}- \kappa_{\mathcal{T}} \Lambda_0 \frac{(e^{\alpha nt-1})^2}{\alpha^2 e^{\alpha nt}}}.
\end{eqnarray}

%%%%%%%%%%%%%%%%%%%%%%%%%%%%%%%%%%%%%%
\begin{figure}
\centering
\includegraphics[width=8cm]{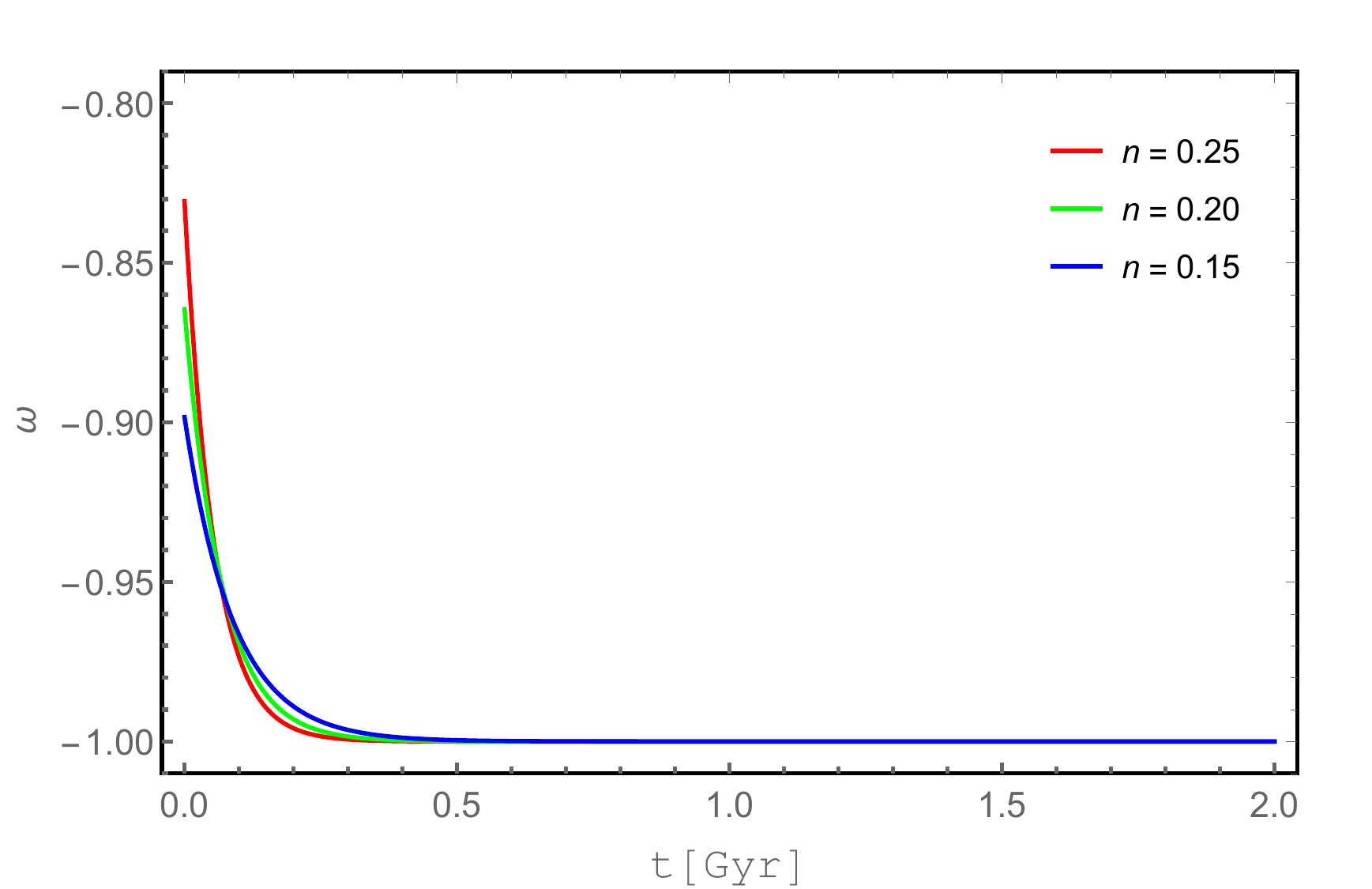}
\caption{Plot for variation of EOS parameter $\omega$ w.r.t. time $t$ (for the specific form of Hubble parameter)}
\end{figure}
%%%%%%%%%%%%%%%%%%%%%%%%%%%%%%%%%%%%%%%

In Fig. 9, the evolutionary aspect of the equation of state parameter for the specific form of the Hubble parameter is shown for three representative values of $n$. Here also, we have considered the coupling constant as $\lambda=-0.5$ and chosen $\Lambda_0=0.1$. The equation of state parameter sharply decreases from a low negative values to overlap with the cosmological constant and becomes $-1$. It is observed that, the evolutionary trajectory of $\omega$ becomes more stiff for a higher value of the parameter $n$.

In this case the null energy condition can be obtained as 
\begin{equation}\label{eq:27a}
p+\rho=\frac{8n(2\pi+\lambda)\alpha^2 e^{\alpha nt}}{\kappa_\mathcal{T}(\kappa_\mathcal{T}+2\lambda)(e^{\alpha nt}-1)^2}.
\end{equation}

The NEC for the specific form of the Hubble parameter is shown in Fig. 10 for both the positive and negative values of $\kappa_\mathcal{T}$. In the model also, a positive value of  $\kappa_\mathcal{T}$ enables the model to violate the null energy condition and for a negative $\kappa_\mathcal{T}$, the model satisfies NEC.

%%%%%%%%%%%%%%%%%%%%%%%%%%%%%%%%%%%%%%%%
\begin{figure}[h!]
\centering
\includegraphics[width=8cm]{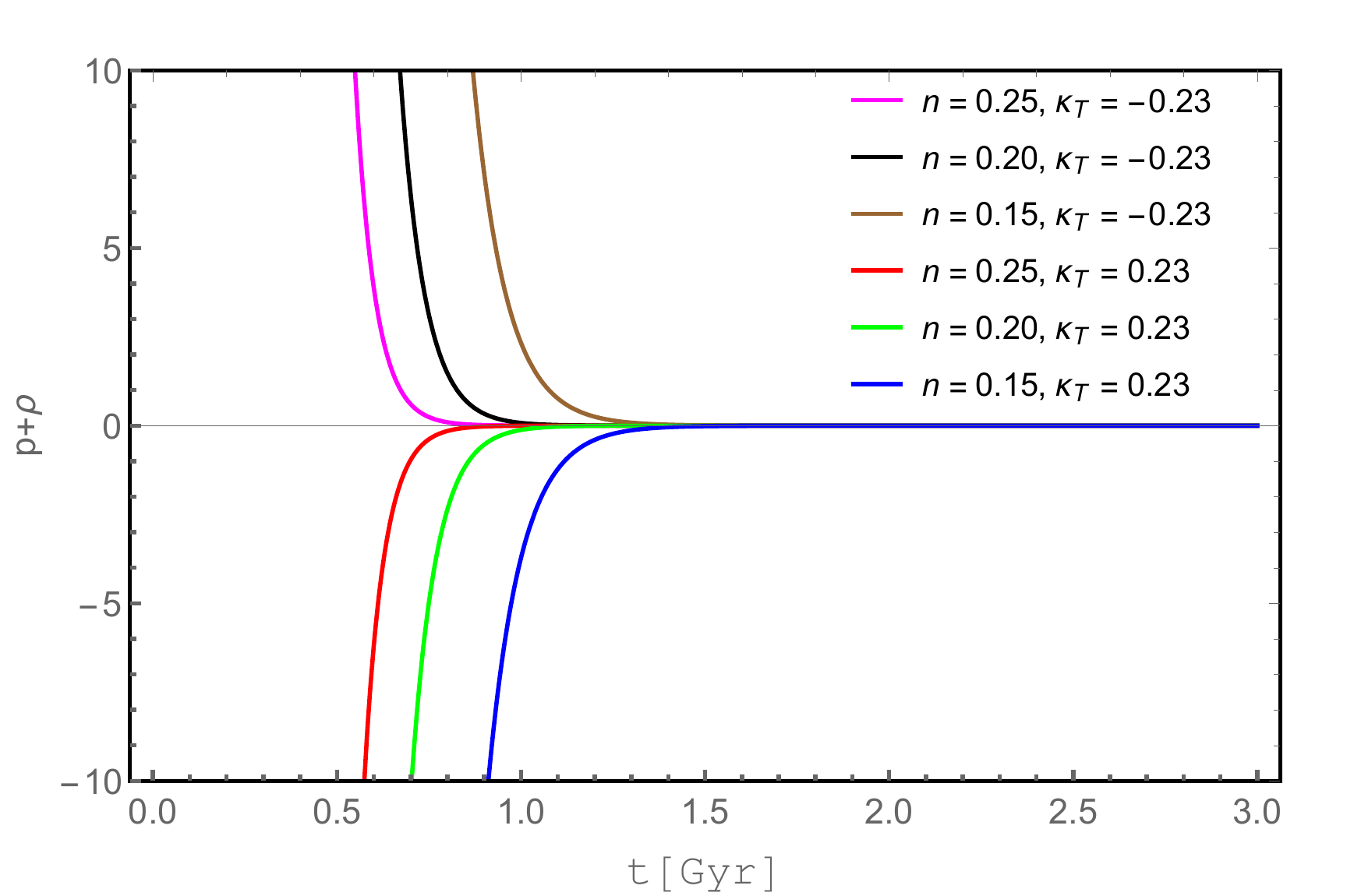}
\caption{p+$\rho$ vs time for the specific form of Hubble parameter. The upper panel is for negative values of $\kappa_\mathcal{T}$ and the lower panel is for positive $\kappa_\mathcal{T}$.}
\end{figure}
%%%%%%%%%%%%%%%%%%%%%%%%%%%%%%%%%%%%%%%%%%%

\section{Wormhole solutions under the $f(R,\mathcal{T})$ gravity}
Let us calculate now the radius of wormhole throat and moreover its evolution due to accretion of phantom energy. It has been argued by Tripathy and Mishra~\cite{Tripathy2020a} that the phantom energy accretion onto wormhole has a definite role to increase the size of the wormhole throat and therefore has a drastic effect to engulf the entire universe before any kind of rip to occur. This type of phenomenon is known as Big Trip in literature under the Big Rip singularity~\cite{Caldwell2003,Nojiri2005b,Frampton2011a,Frampton2011b,Asta2012,Contreras2018,Albarran2018}. From Table 1 it is interesting to note that all the observational data indicate that $\omega <-1$ which refers to the case of phantom model~\cite{Tripathi2017}.

%%%%%%%%%%%%%%%%%%%%%%%%%%%%%%%%%%%%%%%%%%%%%%%%%%%%%%%%%%%
\begin{table*}[htbp!]
\centering 
\caption{The numerical values of the EOS parameter $\omega = p/\rho$ from observational constraints} \label{Table 1}

{\begin{tabular}{@{}ccccc@{}}

$\omega$ & $Observations$ & $Refs.$  \\
      
\hline  ${-1.073}^{+0.090}_{-0.089}$  & WMAP+CMB & \cite{Hinshaw2013} \\ 

\hline  $-1.084 \pm 0.063$ & WMAP+Supernova & \cite{Hinshaw2013} \\ 

\hline  ${-1.035}^{+0.055}_{-0.059}$ & Supernova Cosmology Project & \cite{Amanullah2010} \\ 

\hline  ${-1.06}^{+0.11}_{-0.13}$ & SNLS3+BAO+Planck+WMAP9+WiggleZ & \cite{Kumar2014} \\ 

\hline  $-1.03 \pm 0.03$ & Planck 2018 & \cite{Planck2018} \\ 

\hline  

\end{tabular}}
\end{table*}
%%%%%%%%%%%%%%%%%%%%%%%%%%%%%%%%%%%%%%%%%%%%%%%%%%%%%%%%%%%

Now, by using the evolution equation the Morris-Thorne type wormhole throat radius, $R(t)$, can be estimated for phantom dark energy models~\cite{Babichev2004,Asta2012} as
\begin{equation}\label{eq:28}
\dot{R}= -CR^2(\rho+p),
\end{equation}
where $C$ is a constant which must be positive and dimensionless. 

The integration of the above Eq. (\ref{eq:28}) provides us the wormhole throat radius $R(t)$ as
\begin{equation}
\frac{1}{R(t)}=C\int (p+\rho)dt, \label{eq:28a}
\end{equation}
where we substitute the expressions for $(p+\rho)$ from Eqs. (\ref{eq:24}) and (\ref{eq:27a}) which involves the time dependent factor $\kappa_\mathcal{T}=8\pi+4\lambda \mathcal{T}$. Since $\mathcal{T}=\rho-3p$, $\kappa_\mathcal{T}$ satisfies the equation of the form
\begin{equation}
\kappa_\mathcal{T}^3-a\kappa_\mathcal{T}^2-b\kappa_\mathcal{T}-c(t)=0.\label{eq:28b}
\end{equation}

Here $a=8\pi-2\lambda$ and $b=16\pi\lambda$. The quantity $c(t)$ can be expressed respectively for the model with a Hyperbolic scale factor and the model with a specific form of Hubble parameter as
\begin{eqnarray}
c(t) &=& 12(8\pi+3\lambda)coth^2 (\beta t) -16m(3\pi+\lambda)~cosech^2(\beta t), \label{eq:28c}\\
c(t) &=& 12(8\pi+3\lambda)e^{\alpha nt} -16n(3\pi+\lambda)
\end{eqnarray}

In the above, we consider the cases with vanishing cosmological constant $\Lambda_0$. The solution of Eq. (\ref{eq:28b}) yields
\begin{equation}
\kappa_\mathcal{T}=\frac{a}{3}+\frac{0.26457\left[X-0.41997(-a^2-3b)\right]^{1/3}}{X^{\frac{1}{3}}},\label{eq:28d}
\end{equation}
where $X=2a^3+5.1962Y+9ab+27c(t)$ and $Y=\left[4a^3c(t)-a^2b^2+18abc(t)-4b^3+27c^2(t)\right]^{\frac{1}{2}}$.

Incorporation of such a time dependent $\kappa_\mathcal{T}$ in Eq. (\ref{eq:28a}) makes the things more complicated. In order to obtain possible wormhole solutions with the models discussed in the work, for brevity, we consider constant values for $\kappa_\mathcal{T}$ and denote it as $k$.

\subsection{Model with a Hyperbolic scale factor:}
For the Hyperbolic scale factor (\ref{eq:21}), it is straightforward to obtain from Eq. (\ref{eq:28}) as
\begin{equation}\label{eq:29}
\frac{1}{R(t)}=- \frac{8C(2\pi+\lambda)}{k(k+2\lambda)}\frac{\beta}{m}coth(\beta t)+C_1,
\end{equation}
where $C_1$ is an integration constant. 

At the time of Big Trip ($t_B$), we have $C_1=\frac{8C(2\pi+\lambda)}{k(k+2\lambda)}\frac{\beta}{m}coth(\beta t)$. Consequently, the wormhole radius for the Hyperbolic scale factor case becomes
\begin{equation}
R(t)=\frac{k(k+2\lambda)}{8C(2\pi+\lambda)}\frac{m}{\beta}\left[coth(\beta t_B)-coth (\beta t)\right]^{-1}.
\end{equation}

If the wormhole radius at $t=t_0$ is $R_0$, a Big Trip may occur for the Hyperbolic scale factor model at the epoch
\begin{equation}
t_B=\frac{1}{\beta}coth^{-1}\left[\frac{k(k+2\lambda)}{8C(2\pi+\lambda)}\frac{m}{\beta R_0}+coth(\beta t_0)\right].
\end{equation}

In GR, we have $\lambda=0$ and hence, the epoch of Big Trip becomes
\begin{equation}
t_B=\frac{1}{\beta}coth^{-1}\left[\frac{k^2}{16\pi C}\frac{m}{\beta R_0}+coth(\beta t_0)\right].
\end{equation}

\subsection{Model with a specific form of the Hubble parameter:}
For this case, we get from Eq. (\ref{eq:28})
\begin{equation}
\frac{1}{R(t)}=\frac{8C\alpha(2\pi+\lambda)}{k(k+2\lambda)}\left[\frac{1}{e^{\alpha n t_B}-1}-\frac{1}{e^{\alpha n t}-1}\right].
\end{equation}
Consequently, the throat radius of the Morris-Thorne wormhole for this model becomes
\begin{equation}
R(t)= \frac{k(k+2\lambda)}{8C\alpha(2\pi+\lambda)}
\left[\frac{1}{e^{\alpha n t_B}-1}-\frac{1}{e^{\alpha n t}-1}\right].
\end{equation}

Assuming the throat radius at $t=t_0$ is $R_0$, we obtain the epoch of Big Trip as
\begin{equation}
t_B=\frac{1}{\alpha n}~ln\left[1+\left[\frac{k(k+2\lambda)}{8C\alpha(2\pi+\lambda)R_0}+\frac{1}{e^{\alpha n t_0}-1}\right]^{-1}\right].
\end{equation}

In the limit of GR, we have
\begin{equation}
t_B=\frac{1}{\alpha n}~ln\left[1+\left[\frac{k^2}{16\pi C\alpha R_0}+\frac{1}{e^{\alpha n t_0}-1}\right]^{-1}\right].
\end{equation}

\section{Summary and Conclusion}
Our motivation in the investigation was to present cosmological models under the framework of extended theory of gravity $f(R,\mathcal{T})$. Specifically we have adopted the proposal of Harko et al.~\cite{Harko2011} where $f(R,\mathcal{T})$ has been splited in the form $f(R,\mathcal{T})=R+2f(\mathcal{T})$, considering the Ricci Scalar $R$ and the trace of the energy-momentum tensor $T$ in combination so that effect of both the factors can be obtained in the physical system under consideration. After formulating the basic field equations of the modified gravity theory we have studied two specific cosmological models, firstly the model with Hyperbolic Scale Factor, and secondly the model with specific form of Hubble parameter. These investigations has been followed by finding out of the  wormhole solutions and the possible occurrence of Big Trip in Morris-Thorne wormholes.

We would like to present here some salient features of the models which are as follows:

(i) Figures 1 and 6 represent the Hubble parameters considered in the present work. The parameter space for the forms of the Hubble parameter are adjusted to obtain a reasonable value of the Hubble parameter at the present epoch.

(ii) Figures 2 and 3 present respectively the deceleration parameter and jerk parameter (for case 3.1) whereas Figs. 7 and 8 present respectively the same quantities (for case 3.2) and all of which exhibit expected physical features. One may note that, both the hyperbolic scale factor model and model with the specific form of the Hubble parameter favour an accelerated universe. The choice of the model parameters affect the dynamics of the deceleration parameter and the jerk parameter at an initial epoch.

(iii) Figure 4 shows a unique characteristics of the variation of EOS parameter $\omega$ w.r.t. time $t$. Initially it starts with a bit higher value and then after decreasing follows a straight line showing an ever constant decrement. Thus one can obtain a range of $\omega$ considering the higher and lower part of the plateau which is consistent with the observational values as shown in Table 1. 

(iv) A similar feature as above can be noticed in the model with specific form of Hubble parameter (Fig. 9), however with a rapidly decreasing nature which eventually gets constancy.

(v) In Fig. 5 we have plotted NEC with respect to time for positive and negative $\kappa_T$ for the hyperbolic scale factor model. The figures geometrically show the same feature due to signature flipping in $\kappa_\mathcal{T}$ and thus take the shape of mirror image. The interesting aspect of the graph lies in the pattern which initially decreases sharply and then becomes flat enough and coincides with the zero line. Thus the entire plot is in conformity with the energy condition $p+\rho \geq 0$, i.e. $p+\rho > 0$ (in the upper part) as well as $p+\rho = 0$ (in the lower part) of the first quadrant. However, physically this is not the same for lower quadrant, where the energy condition $p+\rho \geq 0$ does violate for a positive value of $\kappa_\mathcal{T}$. 

(v) The NEC w.r.t. time as plotted for positive and negative $\kappa_\mathcal{T}$ in Fig. 10 display similar behaviour as that in Fig. 5. The model violate the null energy condition for positive $\kappa_\mathcal{T}$.  At this juncture it is interesting to note that the NEC must be violated to make the wormhole remain open and traversable~\cite{MT1988,Rahaman2013,Rahaman2014a,Rahaman2014b,Rahaman2015a,Rahaman2015b}. Therefore, our wormhole solutions as proposed in the present cosmological model is physically viable only for the solution with positive $\kappa_\mathcal{T}$ in Figs. 5 and 10. The solutions with negative $\kappa_\mathcal{T}$ are not admissible.

As a final comment let us add here that physically interesting cosmological models under $f(R,\mathcal{T})$ gravity are possible. Also, it is possible to obtain wormhole solutions in this modified gravity framework as a result of the cosmology violating the null energy condition as supported by the recent observational data.

\section*{Acknowledgement}
Authors are grateful to the Inter-University Center for Astronomy and Astrophysics (IUCAA), Pune, India for hospitality and support during an academic visit where a part of this work is accomplished. The authors are thankful to the anonymous referees for their valuable comments and suggestions for the improvement of the paper.

\end{document}